\documentclass[11pt]{article}
\usepackage[utf8]{inputenc}
\usepackage{longtable}
\usepackage[singlelinecheck=off]{caption}
\usepackage{subcaption}
\usepackage{graphicx}
\usepackage{placeins}
\usepackage{float}
\usepackage{authblk}

\usepackage{amsmath, bbm, amssymb}
\usepackage[dvipsnames,table,xcdraw,svgnames]{xcolor}
\usepackage[hmargin=0.8in,vmargin={1.1in,1.1in}]{geometry}
\usepackage{hyperref}
\usepackage{multicol}

\usepackage{listings}
\hypersetup{
   colorlinks=true,
    linkcolor=blue,
   citecolor=blue,      
   urlcolor=blue,
}
\usepackage{natbib}
\setcitestyle{citesep={;}}
\usepackage[font={footnotesize}]{caption}
\usepackage{url}

\usepackage{parskip}
\usepackage{array}
\newcolumntype{P}[1]{>{\centering\arraybackslash}p{#1}}

\title{Can Generative AI agents behave like humans? \\ 
Evidence from laboratory market experiments}

\author[1,2,3]{R. Maria del Rio-Chanona\textsuperscript{*}}
\author[4]{Marco Pangallo\textsuperscript{*}}
\author[5,6]{Cars Hommes}

\affil[1]{\small{Computer Science Department, University College London}}
\affil[2]{Complexity Science Hub}
\affil[3]{Bennett Institute for Public Policy, University of Cambridge}
\affil[4]{CENTAI Institute}
\affil[5]{Canadian Economic Analysis Department, Bank of Canada}
\affil[6]{Faculty of Economics and Business, University of Amsterdam}
\affil[*]{Equal contribution}

\date{\today}

\begin{document}

\maketitle

\begin{abstract}
We explore the potential of Large Language Models (LLMs) to replicate human behavior in economic market experiments. Compared to previous studies, we focus on dynamic feedback between LLM agents: the decisions of each LLM impact the market price at the current step, and so affect the decisions of the other LLMs at the next step. We compare LLM behavior to market dynamics observed in laboratory settings and assess their alignment with human participants' behavior. Our findings indicate that LLMs do not adhere strictly to rational expectations, displaying instead bounded rationality, similarly to human participants. Providing a minimal context window i.e. memory of three previous time steps, combined with a high variability setting capturing response heterogeneity, allows LLMs to replicate broad trends seen in human experiments, such as the distinction between positive and negative feedback markets. However, differences remain at a granular level--LLMs exhibit less heterogeneity in behavior than humans.  These results suggest that LLMs hold promise as tools for simulating realistic human behavior in economic contexts, though further research is needed to refine their accuracy and increase behavioral diversity.
\end{abstract}

%\tableofcontents

\section{Introduction}

During the 21st century, the academic literature has made significant strides in understanding human behavior in economics, shifting from rational expectations to a more nuanced perspective informed by behavioral economics and experimental research. Despite these advances, challenges remain.. Conducting and replicating experiments that inform these models is resource-intensive, making it difficult to integrate more realistic behavior into economic models with many heterogeneous agents. Recent developments in large language models (LLMs) present a promising solution. Their lower cost compared to traditional laboratory experiments offers a cost-effective alternative for incorporating human behavior into economic modeling  \citep{horton2023large}. However, before LLMs can be used to model economic behavior accurately,  we must evaluate their capabilities in replicating both individual behaviors and market interactions. In this paper, we explore the extent to which LLMs can replicate human behavior in laboratory market settings and identify conditions under which they exhibit human-like economic behaviour.

% LLMs in social science

LLMs’ ability to simulate human interactions has led academics to suggest they could transform social sciences \citep{grossmann2023ai}. LLM agents in social systems have demonstrated human-like characteristics and produced emergent phenomena, such as organizing events or making decisions about pandemic response \citep{park2023generative, williams2023epidemic}. They have also passed the Turing test in various games and experiments \citep{mei2024turing} and shown heterogeneity (e.g. \cite{argyle2023out} found that LLMs could adopt political identities, such as Republican or Democrat, depending on the prompts given).

While these studies show that LLMs can simulate social interactions, modeling economic behavior poses different challenges due to its reliance on quantitative reasoning. Yet, research shows that LLMs can reproduce to some extent the economic behaviour of individuals. \cite{horton2023large} showed that LLM agents can behave similarly to humans when it comes to status quo bias and labour substitution with respect to minimum wage, and \cite{mei2024turing} shows that LLMs pass the Turing test in economic games such as public goods, the ultimatum game, and the prisoners' dilemma. When faced with budgetary decisions, LLMs exhibit rational behaviour \citep{chen2023emergence} that is somewhat similar, though not perfectly aligned, with humans with respect to risk, social, food, and time preferences. 

These studies focus on individual economic behaviors, but much less existing work aims at understanding how the interaction of multiple LLMs plays out \textit{over time}. For example, would individual agents interacting in a market converge to the rational equilibrium price? Even if they converge, what type of fluctuations would they produce on their way to equilibrium? Although a few studies have looked at dynamic settings, the temporal dimension is not explored in detail. For instance, \cite{chen2023put} simulate auction markets, but mostly focus on the relative effectiveness of different bidding strategies. Similarly, \cite{fish2024algorithmic} examined firms’ pricing strategies in an oligopoly setting, but only focus on the final prices firms settle on.\footnote{Moreover, \cite{zhaocompeteai} simulated an economic environment involving restaurants and consumers, and find the emergence of diverse strategies from restaurants to attract customers. \cite{li2024econagent} study a macroeconomic agent-based model, focusing on the replication of stylized facts.}

Moreover, these studies do not compare LLM-generated dynamics to empirical or experimental dynamics.\footnote{Some studies achieved partial validation through static stylized facts; for example, \cite{zhaocompeteai} reproduced the Matthew effect, and \cite{li2024econagent} captured Okun's law, but failed to capture the Phillips curve.} To fully understand the potential of LLMs in economic contexts, we must explore their capacity to replicate market dynamics and agent interactions while validating these simulations against detailed human behavior. This is particularly challenging for three reasons: First, modeling market dynamics requires LLM agents to understand money, quantities, and mathematical concepts—areas where they initially struggled. Second, unlike single economic experiments, simulating markets hinges on interaction. This requires effectively translating these interactions into a virtual environment and providing LLM agents with the necessary context to engage iteratively without exhausting context window limits. Third, establishing a ground truth for behavior in markets is difficult since experiments of humans in markets are scarcer than those of individual economic behaviour. 

To address these challenges, we focus on market dynamics well-documented in human experiments, particularly the literature on price expectations and the behavior of participants in laboratory market experiments \citep{heemeijer2009price}\footnote{These market experiments are in fact repeated Keynes' beauty contest or guessing games as introduced in \citep{Nagel1995}; See \citep{MauerNagel2018} for a survey of beauty contest games in the lab.}{}. This body of work has been crucial in understanding how expectations form and deviate from rational expectations \citep{hommes2021behavioral,anufriev2012evolutionary}.  Market participants use simple heurisitcs\footnote{In the psychology literature, heuristics have been studied extensively in the last few decades, following the seminal work of 
\cite{Tversky1974} and \cite{Gigerenzer2011}}
—fundamentalists, trend followers, naive forecasters, and adaptive learners—each influence market outcomes in distinct ways. An interesting finding is that market behavior also varies by market type. In positive feedback markets, dominated by trend followers, self-reinforcing behaviors often lead to bubbles \citep{hommes2008expectations}. In contrast, negative feedback markets, where fundamentalists are prevalent, tend to be more stable, with prices closely aligning with fundamentals \citep{heemeijer2009price}.

We choose this literature due to its relevance in behavioural economics \citep{hommes2021behavioral} and for two practical reasons. First, the experiments were conducted with written instructions, and participants interacted through a computer, making it easier to translate the experiment into an LLM environment. Second, these experiments are among the few that capture detailed market dynamics and participant interactions, allowing for a thorough comparison.

In this paper, we contribute to the literature by providing a framework for translating laboratory experiments into LLM experiments, documenting the roles of key parameters such as context window (memory), response variability (known technically as \textit{temperature}), and model type. Our main finding is that, under a context window of at least 3 time steps and high response variability, GPT 3.5 and GPT 4 LLM markets exhibit differences between positive and negative feedback markets similar to those observed in human experiments. Specifically, LLMs demonstrate rapid, oscillating convergence in negative feedback markets and slower convergence in positive feedback markets toward the equilibrium price. Additionally, when analyzing LLM strategies, we find less heterogeneity compared to human participants. Importantly, LLM markets do not adhere strictly to rational expectations but instead exhibit bounded rationality. 

Overall, our results suggest that LLMs are a promising tool for simulating realistic human behavior beyond rational expectations. However, this potential depends significantly on the parameters used. We recommend a high response variability and a context window of at least 3 time steps. To achieve the ultimate goal of using LLMs to simulate economic behaviour, significant studies should be done to document where LLMs aligns and where they do not.

The paper is structured as follows. Section~\ref{sec:model} introduces the experimental market framework and the feedback mechanisms, how we translate the experimental design to an LLM environment, and how we measure alignment between human and LLM agents. Section~\ref{sec:results} reports the main results, comparing market dynamics and forecasting behavior. Section~\ref{sec:discussion} discusses implications and limitations.

\section{Methods}
\label{sec:model}
\subsection{Original experiments}
We simulate a laboratory experiment based on \cite{heemeijer2009price}, where groups of six individuals attempt to predict the price of a good in the next time step for a total of 50 time steps. The realized price at each step is determined mostly by the aggregate predictions of the six participants, though a small amount of noise is also introduced. Participants do not have direct contact with one another; instead, they interact indirectly through the market information displayed on their computer screens. The motivation behind this experiment is that a market, like other social environments, can be viewed as an expectations feedback system. In such systems, past market behavior influences individual expectations, which in turn determine current market behavior. This feedback loop creates a dynamic environment where expectations and market outcomes are interdependent. 

\subsubsection{Market Dynamics}
\label{sec:markey_dynamics}
There are two types of markets in the \cite{heemeijer2009price} experiments: one characterized by positive feedback loops and another by negative feedback loops. 

In markets with positive feedback, a higher average price forecast by participants leads to an increase in the realized market price. This type of feedback resembles speculative financial markets, such as stock or real estate markets, where the expectation of rising prices can lead to increased demand, further driving up prices. These dynamics was captured, among others, in the asset pricing model by \cite{brock1998heterogeneous}. In this model, traders choose between a risk-free and a risky asset, deciding their demand of the risky asset based on their expectations of its price. Excess demand
for the risky asset leads to an increase in the asset price. Following a version of the model with a market maker \citep{hommes2005robust}, which does not require to explictly compute equilibrium \citep{pangallo2024equations}, and selecting specific parameter values, \cite{heemeijer2009price}  obtain a very simple price adjustment rule for the positive feedback market:
\begin{equation}
    p_t = \frac{20}{21}\left(\bar{p}^e_t + 3\right) + \epsilon_t,
    \label{eq:positive_market}
\end{equation}
where \(p_t\) is the realized market price at time \(t\), \(\bar{p}^e_t\) is the average forecast of the six participants, and \(\epsilon_t \sim N(0, \frac{1}{4})\) is a normally distributed random term representing small random fluctuations. In this scenario, the market price increases when participants collectively anticipate higher future prices.

In contrast, in markets with negative feedback, a higher average price forecast leads to a decrease in the realized market price. This type of feedback is often seen in markets for non-storable goods, such as agricultural products, where higher expected future prices lead to increased current production, which in turn results in lower realized prices. These dynamics are exemplified in the so-called cobweb model \citep{ezekiel1938cobweb}. In the model, demand is exogenous and depends negatively on price, while supply is endogenous and is set to maximize profits depending on firms' expectations of future prices. In the version of the cobweb model used by \cite{heemeijer2009price}, markets do not clear ex ante. Instead, prices adjust in the direction of excess demand. Choosing appropriate parameters for the demand and supply curves, the price adjustment rule for the negative feedback market is
\begin{equation}
    p_t = \frac{20}{21}\left(123 - \bar{p}^e_t\right) + \epsilon_t.
\label{eq:negative_market}
\end{equation}
where, as before, \(p_t\) is the realized market price at time \(t\), \(\bar{p}^e_t\) is the average forecast, and \(\epsilon_t \sim N(0, \frac{1}{4})\) is a normally distributed random term.

In both market types, the rational equilibrium price \(p^* = 60\) represents a steady-state condition where the market price would remain stable if all participants predicted this price. In the real world, this equilibrium price can be interpreted as the fundamental value of an asset or the long-term average price in a market without external shocks or speculation. The price of 60, therefore, serves as a benchmark around which participants' forecasts and actual prices may fluctuate. The absolute value of the slope \( \frac{20}{21} \) is less than one, ensuring stability under naive or adaptive expectations, but close enough to one to result in relatively slow convergence, thereby allowing us to observe the dynamics of the participants' expectations over time.

\subsubsection{Human Participant Interaction}
Participants were asked to forecast the price of the good for the next period in each of the 50 rounds of the experiment. Their earnings depended on the accuracy of their predictions, incentivizing them to make precise forecasts. Specifically, the earnings \(E_h\) for participant \(h\) in period \(t\) were determined by the following quadratic loss function
\begin{equation}
    E_h = \max \left(1300 - \frac{1300}{49} \left(p_t - \hat{p}_{h,t}\right)^2, 0\right),
    \label{eq:earnings}
\end{equation}
where \(p_t\) is the realized market price, and \(\hat{p}_{h,t}\) is the forecast submitted by participant \(h\) for period \(t\). Participants were provided with information about past realized prices and their own past predictions, but not with the predictions of others, simulating the limited information conditions often present in real-world markets.

\subsection{Translating the experiment to Large Language Model's environment}
We translate the experiments by \cite{heemeijer2009price} into an LLM environment through the OpenAI API. To translate the environment settings to LLMs we make use only of the text capabilities via the API and simulate 50 time steps iteratively as in the experiment. We briefly outline the parameters and how LLMs API works below.

\subsubsection{OpenAI API, Models, Temperature, and  Memory}
\paragraph{Model.} The core of the OpenAI API is the model, like GPT-4 or GPT-3.5. These are pre-trained neural networks that process and generate human-like text based on input they receive. When you send a request to the API, the model processes the input (a "prompt") and generates a response. 
\paragraph{Temperature.}
The temperature parameter controls the randomness of model output. It is typically a value between 0 and 1, though it can exceed 1. Lower temperatures (closer to 0) make responses more deterministic. Higher temperatures introduce more randomness, allowing for more diverse outputs but potentially reducing coherence.
\paragraph{Messages, Roles, and context window.}
The OpenAI API uses a chat-based format where messages are exchanged between different "roles": system, user, and assistant. System messages set the behavior of the assistant and usually prime the model. User messages represent input from the person using the model. Assistant messages represent the output generated by the model.
These messages are processed in sequence to build context. The context window is the total amount of text (in \textit{tokens}, roughly corresponding to characters) the model can consider at once. If a model has a 4,096-token context window and the conversation exceeds this limit, earlier messages might be "forgotten" as newer ones take priority.

\paragraph{Memory.}
For this research, we define memory as the number of previous messages fed into the system when generating a response. Memory determines how much conversation history the model can access, influencing the coherence of its output. Memory 0 provides no prior context, while memory 1 includes only the last interaction. In our 50-step experiments, the maximum possible memory would be 50, allowing the LLM agent to retain information from each step.
LLM agents are always provided with the full time series of previous predicted and realized prices. Memory determines whether the agent retains the reasoning behind its previous predictions.

\paragraph{Seed.}
A seed initializes the random number generator influencing model outputs. In the OpenAI API, seeds improve reproducibility but don't guarantee perfect determinism across all conditions or API versions. We use seeds to enhance experimental consistency.

We explore the behaviour of LLM-agent markets under positive and negative feedback loops, testing all combinations of memory 1, 3, and 5 and temperature 0.3, 0.7, and 1.0. We rely on both gpt-3-5-turbo-1106 (from now on, GPT 3.5) and gpt-4-1106-preview (GPT 4). We queried the API between March and April 2024.

\subsubsection{Prompts and setting of the LLM experiments}

In translating the experimental setup by \cite{heemeijer2009price} into a Large Language Model (LLM) environment, we keep the instructions between GPT and human participants as close as possible. However, we add at the end instructions on how the text format through which market information will be given and the format in which they should supply the answer.

In designing the prompts and the format in which LLMs should provide their output, we follow two key principles\footnote{\url{https://learn.deeplearning.ai/chatgpt-prompt-eng}}: (i) provide clear and specific instructions, and (ii) allow the LLMs 'time to think,' also known as the chain-of-thought technique. To align with the first principle, we use delimiters and clear punctuation to delineate different sections of the instructions. Additionally, we request that the output be provided in JSON format via the OpenAI API, which facilitates structured interactions. The original instructions given to human participants were adapted to include these additional guidelines. For the second principle, we prompt the LLM to give its reasoning before providing its prediction. (Appendix \ref{apx:instructions} reports the instructions given to human participants in \cite{heemeijer2009price}, while Appendix \ref{apx:llm_instructions} details the prompts given to the LLM.)

As in the original experiments by \cite{heemeijer2009price}, we simulate markets with six agents over 50 time steps. LLM agents are instructed that their goal is to maximize profits, which are determined by the accuracy of their predictions. They are also informed about whether the market has positive or negative feedback dynamics, with a description of the market similar to the one in Section \ref{sec:markey_dynamics}.  At the initial time step, the LLM agents receive instructions informing them that no prior market information is available, and they must guess an initial price between 1 and 100. The simulation proceeds iteratively as follows:

\begin{enumerate}
    \item \textbf{Instructions Delivery:} At the start of each time step, LLM agents receive the experiment's instructions, including a reminder of their role as advisors, type of feedback market, goal to maximize earnings and that their earnings depend on the accuracy of their predictions.
    \item \textbf{Context and Memory:} LLM agents are provided with the context corresponding to the memory parameter \(m\). This context includes the reasoning and predictions from previous time steps as determined by the memory setting. For example, with memory 1, the LLM recalls the reasoning and prediction from the immediately preceding time step.
    
    \item \textbf{Market Information:} Agents receive current market information, which includes the realized market price from the previous time step, historical market prices, and their current accumulated earnings.
    
    \item \textbf{Prediction Task:} LLM agents are asked to first provide their reasoning and then predict the next period's price. 
    
    \item \textbf{Market Price and Earnings Calculation:} Once all LLM agents have submitted their predictions, we calculate the market price for the current time step using Eq. \ref{eq:positive_market} or \ref{eq:negative_market} depending on the market type. We compute the earnings for each agent using Eq. \ref{eq:earnings}. 
\end{enumerate}
This iterative process allows us to observe how LLM agent markets behave and how this behavior depends on the model, memory, temperature and type of market.

\subsection{Measuring Alignment}
\label{sec:alignment}

To measure the alignment between the LLM-simulated experiments and the human participants, we estimate a regression of how predictions depend on past prices and past predictions. \cite{heemeijer2009price} and \cite{anufriev2012evolutionary} find that a very simple rule that only uses one lag of past prices and past predictions explains the behavior of a large number of experimental participants. According to this \textit{first-order heuristic}, the price expectation of agent $h$ at time $t$ is given by
\begin{equation}
    p^e_{h,t} = \alpha_1 p_{t-1} + \alpha_2 p^e_{h,t-1} + (1 - \alpha_1 - \alpha_2) \cdot 60 + \beta \cdot (p_{t-1} - p_{t-2}) + \nu_t.
    \label{eq:firstorder}
\end{equation}
Here, \(\alpha_1\), \(\alpha_2\), and \(\beta\) range between -1 and 1, representing the strategies of each agent. To illustrate the meaning of these parameters, it is useful to consider a few special cases.
\begin{itemize}
    \item Fundamentalism ($\alpha_1 = 0$, $\alpha_2 = 0$, $\beta_1 = 0$). In this case, up to noise the agent always expects the fundamental price (60).\footnote{Note that fundamentalism is not the same as rational expectations, as rational expectations agents take the behavior of other agents into account, and adjust their strategy to maximize their profits. By contrast, a fundamentalist always expects the fundamental price no matter what the other agents are going to do.} 
    \item Naivety ($\alpha_1 = 1$, $\alpha_2 = 0$, $\beta_1 = 0$). In this case, the agent just relies on the last observed price, displaying naive expectations.
    \item Obstinacy ($\alpha_1 = 0$, $\alpha_2 = 1$, $\beta_1 = 0$). In this case, the agent keeps believing on its past expectation, however that performed.
    \item Trend following ($\alpha_1 = 0$, $\alpha_2 = 0$, $\beta_1 = 1$). Here, the agent expects trends to persist: if prices have been going up leading to the current step, he expects that prices will keep going up. 
    \item Trend reversing ($\alpha_1 = 0$, $\alpha_2 = 0$, $\beta_1 = -1$). Here, the agent expects trends to revert: if prices have been going up leading to the current step, he expects that prices will go down.
\end{itemize}
Aside from these extreme cases, any parameter combination can describe the extent to which agents rely on one forecasting rule or another. For instance, we can label ``adaptation'' a situation in which $\alpha_1=0.5$ and $\alpha_2=0.5$, that is agents partially rely on past prices, while slowly updating their expectations.

We estimate Eq. \eqref{eq:firstorder} using Ordinary Least Squares (OLS), treating each human participant or LLM agent separately. Following \cite{heemeijer2009price}, we follow an iterative procedure to drop coefficients that are not significant at the 5\% level. More specifically, we start by doing OLS estimation on the full equation, and then, if some coefficient is not significant, we drop the one with highest p-value, performing again the OLS estimation on the reduced equation. We repeat this operation until all remaining coefficients are statistically significant. We also follow  \cite{heemeijer2009price} in removing an initial learning phase in which at most three out of six participants are within 5\% of the market price, estimating parameters once market dynamics are more settled.\footnote{We also drop a number of anomalous data points in the experiments with human subjects, likely reflecting errors by the participants. See \cite{heemeijer2009price}.} However, we also consider robustness to this assumption, estimating parameters on the entire time series, too.

We can in principle check how the behavior of individual LLM agents compares to human participants. However, to be more systematic and compare different hyper-parameters such as memory and temperature, we also rely on aggregate metrics. Because trend following behavior seems to be the most important factor that determines how human participants behave in positive vs. negative markets, we assess alignment between human participants and LLM agents by averaging the \(\beta\) coefficient across agents. In the experimental positive feedback markets of \cite{heemeijer2009price}, this coefficient is high and positive (0.67), suggesting agents follow trends. In contrast,  in negative feedback markets this coefficient is zero, suggesting that agents ignore trends. Do LLM agents behave similarly?

\section{Results}
\label{sec:results}

\subsection{Comparing market dynamics}
\label{sec:results_market_dynamics}

\begin{figure}[H]
    \centering
    \includegraphics[width=1\textwidth]{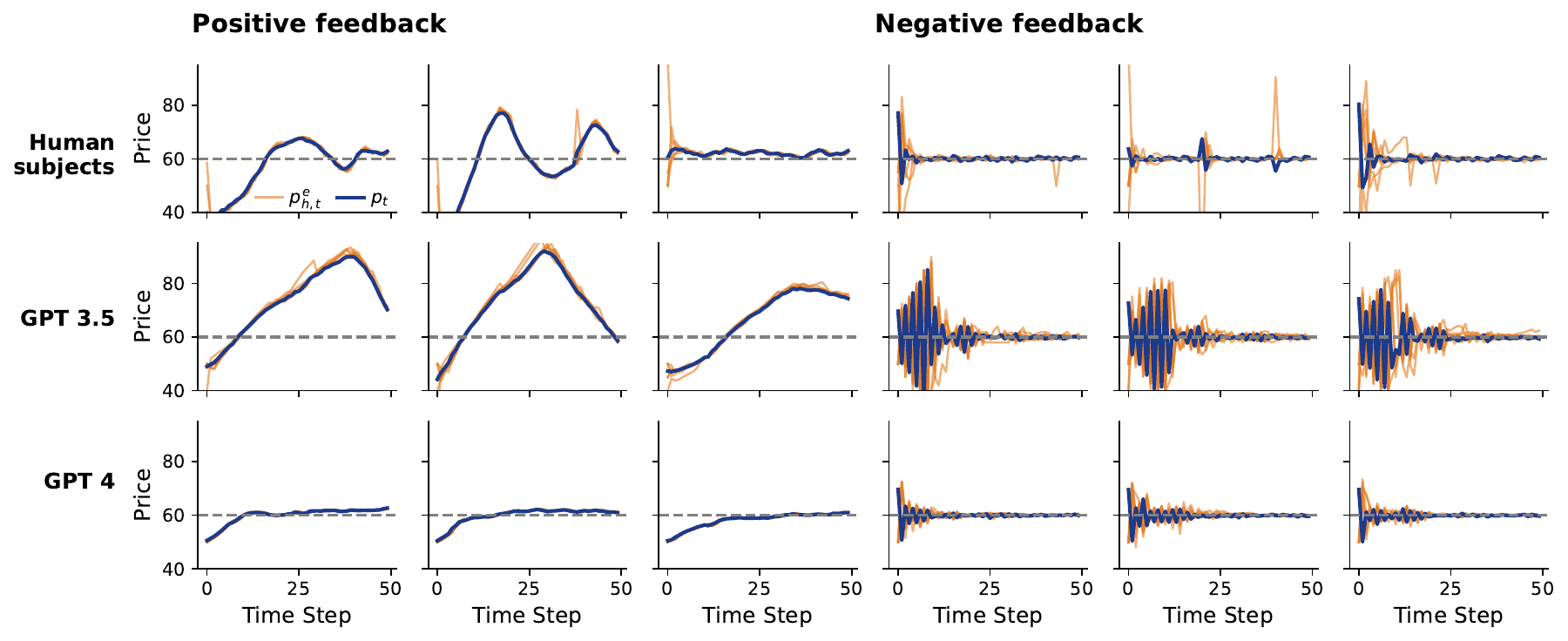}
    \caption{\textbf{Market dynamics for human subjects and LLM agents.} We compare three different experiments for both positive and negative feedback markets. For each experiment, we show both the time series of realized market price (blue) and all agents' expectations (orange). For this illustration of LLM behavior, we select a specific combination of memory (3) and temperature (1.0) that seems to best capture experimental market dynamics.}
    \label{fig:grid_human_ai}
\end{figure}

We compare the market dynamics of human and LLM-based agents by analyzing their expected and realized prices. Figure \ref{fig:grid_human_ai} presents the time series for human subjects and a selected set of GPT-3.5 and GPT-4 parameters. For both negative and positive feedback markets, we show results from three independent experiments. Supplementary Figure \ref{fig:Gridplot_human_models_supp} confirms that other experiments and parameter settings yield qualitatively similar results.

\paragraph{Positive feedback markets}
In positive feedback markets, human subjects tend to fluctuate significantly around the equilibrium price. In the top-left panel, after substantial fluctuations, they appear to reach equilibrium by the end of the experiment. However, in the other two cases, they either continue fluctuating without stabilizing or maintain smaller oscillations slightly above the equilibrium price. GPT-3.5 agents also exhibit large fluctuations, though at a slower pace, and fail to converge to equilibrium—--similarly to the second experiment with human subjects. In contrast, GPT-4 agents quickly stabilize at a steady-state price slightly above equilibrium, resembling the third human experiment.

The cases with large fluctuations are particularly interesting, as both human subjects and GPT-3.5 agents display bubble-like behavior. Notably, the underlying market model is linear, meaning there is no built-in turning point in the equations governing its dynamics. A positive trend can, in theory, persist indefinitely, with prices rising arbitrarily high as long as agents collectively predict continued increases. Yet, despite this, market dynamics eventually reverse. Interestingly, GPT-3.5 agents take much longer to revert the trend compared to human participants. We explore this phenomenon further in Section \ref{sec:results_textual}, analyzing the narratives provided by GPT-3.5.

\paragraph{Negative feedback markets}
In negative feedback markets, human participants typically converge to the equilibrium price within ten time steps. Large oscillations (exceeding 20 EUR from equilibrium) are mostly confined to the first few time steps, after which stability is achieved. Occasionally, an individual participant makes a forecast far from equilibrium, but this alone is insufficient to shift market prices significantly. GPT-3.5 agents, however, behave quite differently. Prices and expectations oscillate repeatedly, requiring approximately 25 time steps to converge. GPT-4, in contrast, exhibits behavior more similar to humans, reaching equilibrium in about 10–15 steps (see Figure \ref{fig:grid_human_ai} right panels).

These results suggest that LLM agents exhibit a degree of naïvety in negative feedback markets. Because they expect prices to remain close to previous values, their forecasts create supply-demand imbalances---excess supply when predicted prices are too high and excess demand when they are too low. As a result, market prices oscillate around equilibrium, gradually stabilizing. However, particularly in the case of GPT-3.5, this convergence is not monotonic. At times, the amplitude of oscillations increases before eventually declining.

\begin{figure}[H]
    \centering
    \includegraphics[width=1\textwidth]{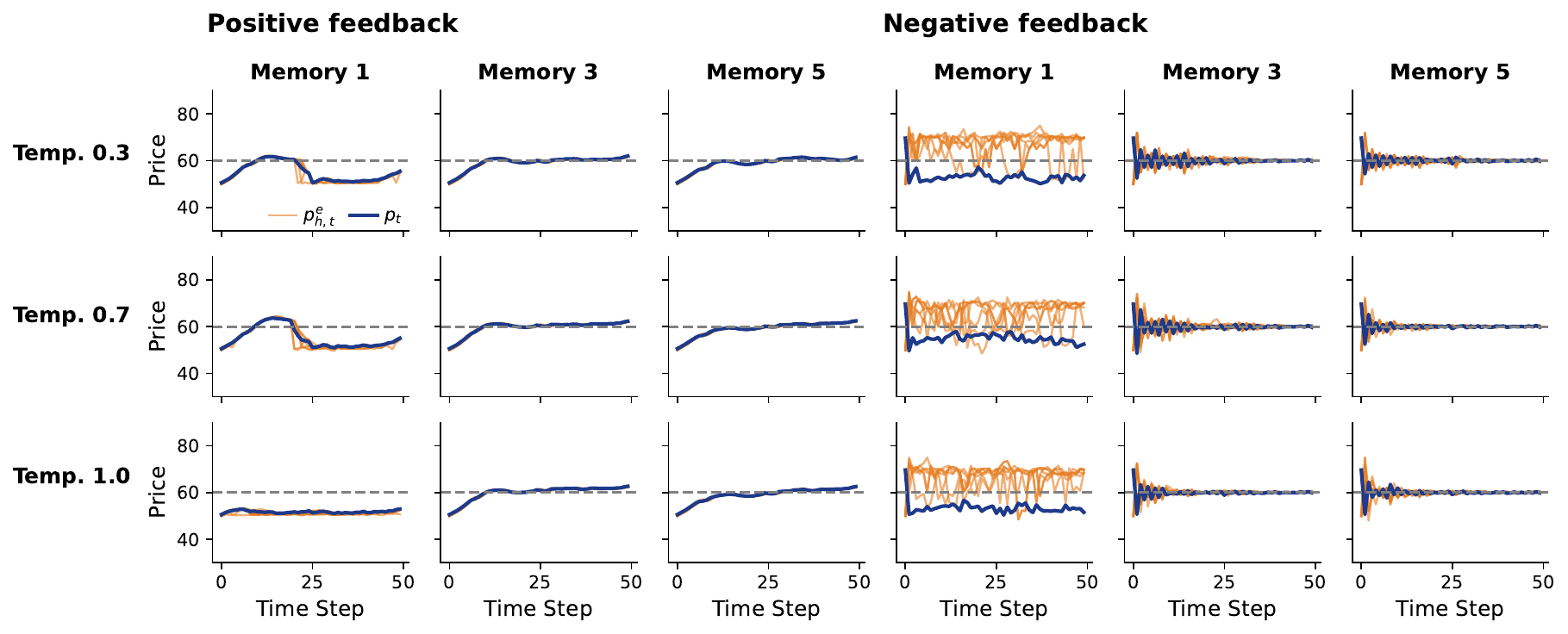}
    \caption{\textbf{Market dynamics for GPT-4 agents under different parameters.} We compare all nine combinations of memory (1, 3, 5) and temperature (0.3, 0.7, 1.0), for both positive and negative feedback markets. For each experiment, we show both the time series of realized market price (blue) and all agents' expectations (orange).}
    \label{fig:grid_gpt4}
\end{figure}

\paragraph{Memory and temperature effects}
How robust are these results across different values of memory and temperature? In Figure \ref{fig:grid_gpt4} we show nine combinations of memory and temperature, for GPT-4 agents in negative and positive feedback markets. (Supplementary Figure \ref{fig:grid_gpt3} shows results for GPT-3.5 agents.)

The market dynamics of LLM agents depend strongly on memory and, to a lesser degree, on temperature. In positive feedback markets, GPT-4 converges quickly when memory is set to 3 or higher, but it fails to converge with memory 1. A similar pattern emerges in negative feedback markets: with memory 3 or 5, GPT-4 mirrors human behavior, converging within 15 time steps even at low temperature, but it does not converge at memory 1. Overall, these findings highlight memory as a key parameter for producing human-like market behavior; while higher temperatures also appear beneficial, the effect of memory is dominant.

Overall, our results highlight both similarities and differences between human subjects and LLM-based agents in market dynamics. In positive feedback markets, both humans and GPT-3.5 agents exhibit large fluctuations, sometimes stabilizing near equilibrium and sometimes continuing to oscillate. GPT-4 agents, like some human participants, tend to converge more reliably to a steady-state price slightly above equilibrium. In negative feedback markets, all groups ultimately reach equilibrium, but the speed and path differ: humans typically stabilize within ten time steps, while GPT-3.5 agents take around 25 steps, and GPT-4 behaves more similarly to humans, stabilizing within 10–15 steps. A key similarity is that both humans and LLM agents show trend-following behavior in positive feedback markets but do not exhibit clear trend-following in negative feedback markets. Importantly, the market behavior of LLM agents is highly sensitive to memory, with convergence occurring only when memory is set to 3 or higher, highlighting its role as a key parameter for human-like performance.

\subsection{Estimating behavioral parameters}
\label{sec:results_behavioral_parameters}

\begin{figure}[!h]
    \centering
    \includegraphics[width=1\textwidth]{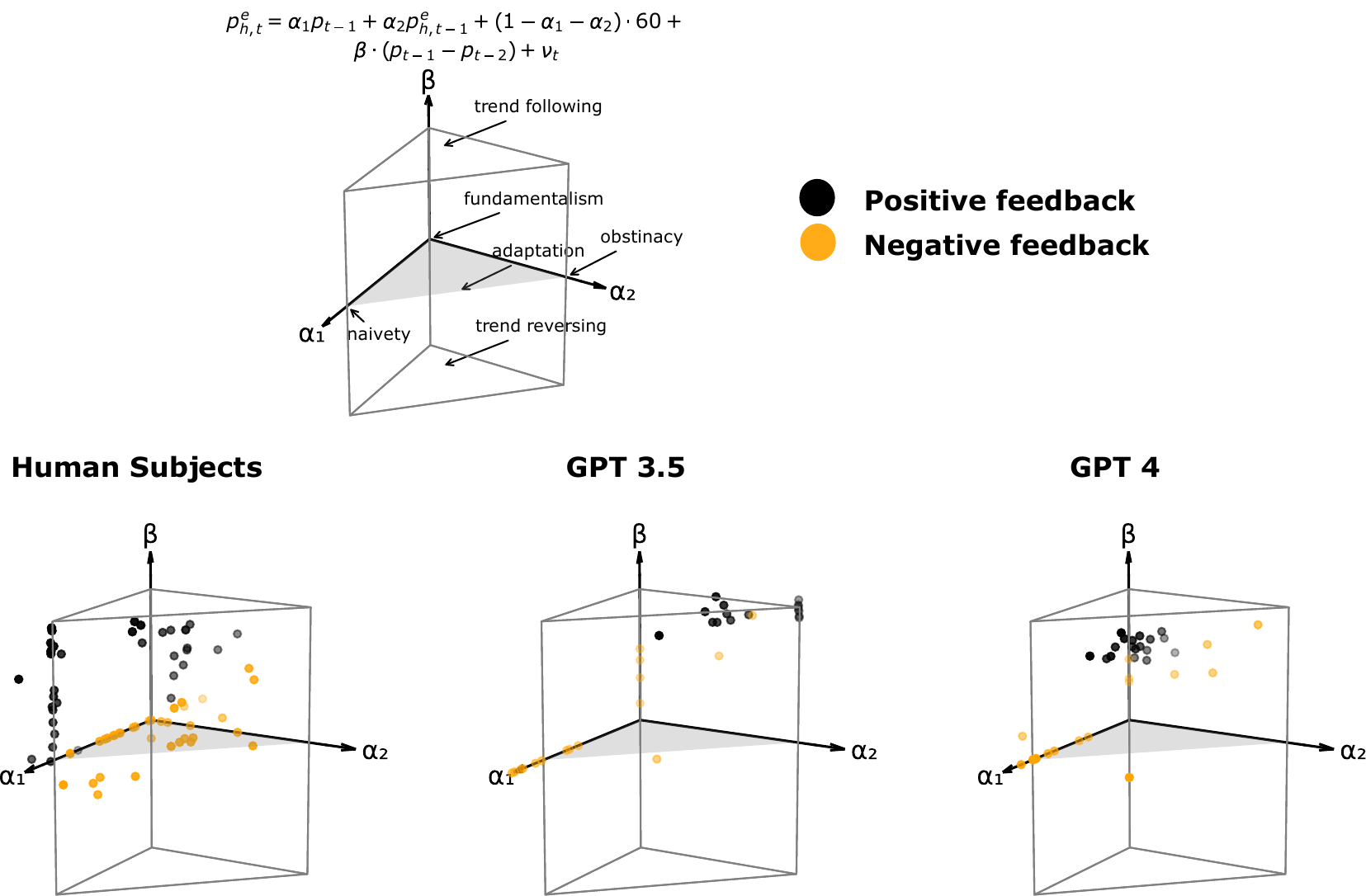}
    \caption{Comparison of strategies between human subjects and LLM agents. Each prism of first-order heuristics shows the position of the point $(\alpha_1, \alpha_2, \beta)$. Positive (negative) values of \( \beta \) indicate trend-following (trend-reversing) prediction rules, naivety corresponds to $(1,0,0)$, obstinacy to $(0,1,0)$ and fundamentalism to $(0,0,0)$, while adaptation is a situation with $\alpha_1, \alpha_2 >0$, such as $(0.5,0.5,0)$. In the prisms at the bottom, we show the estimated parameters for each human subject or LLM agent.  Orange dots represent estimated parameters in negative feedback markets, while black dots represent parameters from positive feedback markets. As in Figure \ref{fig:grid_human_ai}, we focus on memory 3 and temperature 1.0.}
    \label{fig:prism}
\end{figure}

To perform a more quantitative comparison between human subjects and LLM agents, we estimate the parameters of the first-order heuristic following the procedure described in Section \ref{sec:alignment}, focusing on memory 3 and temperature 1.0 for illustration. Figure \ref{fig:prism} presents the estimated strategy profiles of individual forecasters as points in a three-dimensional space defined by $\alpha_1$, $\alpha_2$, and $\beta$.  

Human subjects exhibit greater heterogeneity than LLM agents, with their estimated parameters more widely dispersed across the parameter space. In positive feedback markets, all human participants appear to be trend followers ($\beta > 0$), but they separate into two distinct clusters: one of ``pure trend-followers'' ($\alpha_1 = 1$), using the last price as an anchor, and another of adaptive forecasters. In contrast, no clear structure emerges in negative feedback markets, where human subjects do not exhibit systematic trend-following behavior ($\beta \approx 0$).  

GPT-3.5 and GPT-4 agents show similar trend-following behavior. In both cases, LLM agents act as trend followers in positive feedback markets but do not appear to follow trends in negative feedback markets. However, their behavior differs from that of human subjects in other respects. In positive feedback markets, LLM agents lack a distinct cluster of naïve forecasters, whereas in negative feedback markets, all agents appear to be naïve.  

\begin{figure}[H]
    \centering
    \includegraphics[width=1\textwidth]{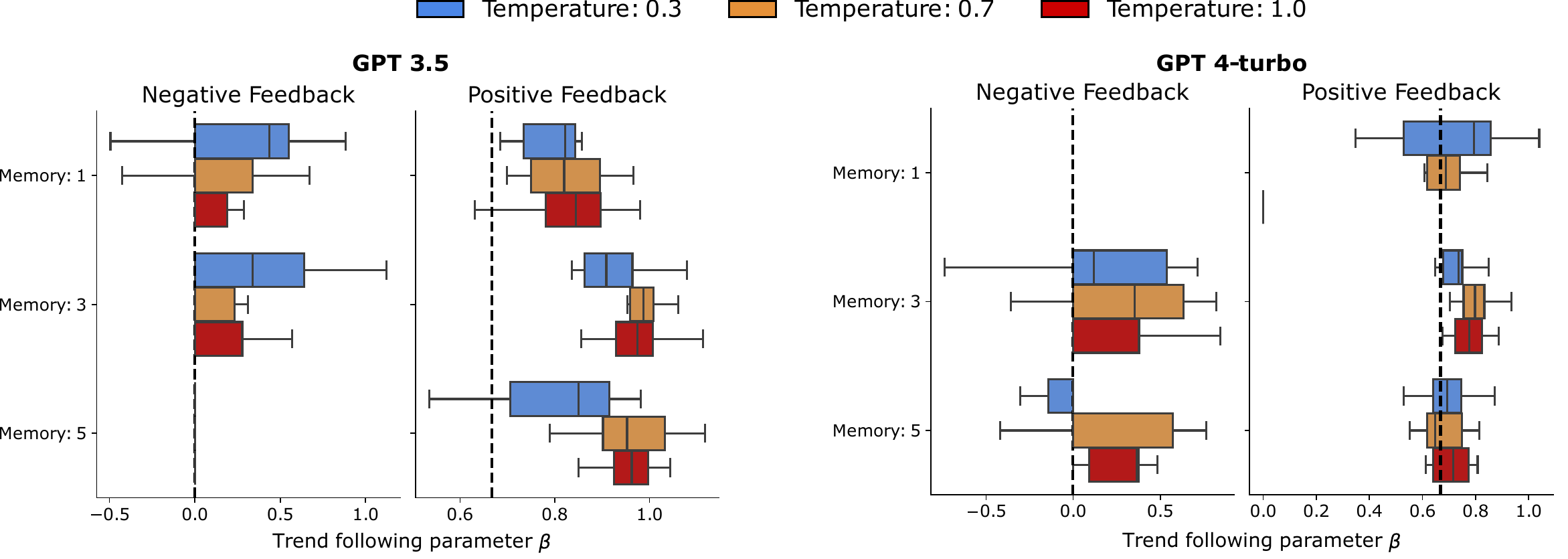}
    \caption{\textbf{Trend following behavior across different memory and temperature parameters.} Each box-plot shows the median, interquartile range, and whiskers of all the estimated $\beta$ values for LLM agents. Dashed vertical lines indicate the average $\beta$ values for human participants, namely 0 for negative feedback markets and 0.67 for positive feedback markets. GPT-3.5 with memory 5 and negative feedback always has $\beta=0$, so the box-plots only show as interruptions of the dashed line. GPT-4 with memory 1 and negative feedback cannot be used to estimate the $\beta$ parameters, and so it does not show any result (see text).}
    \label{fig:table_overview}
\end{figure}

To systematically investigate the estimated strategies across different memory and temperature parameters, we focus on trend following. Figure \ref{fig:table_overview} presents box plots of all estimated $\beta$ parameters across LLM agents for various combinations of memory and temperature.  

Overall, the main finding is robust for both GPT-3.5 and GPT-4: in negative feedback markets, $\beta$ remains close to zero, while in positive feedback markets, it is consistently positive. However, differences emerge across memory and temperature settings, as well as between GPT-3.5 and GPT-4. Under negative feedback, GPT-3.5 with low temperature results in a median $\beta$ that is slightly positive, although with memory 5, all estimated $\beta$ values are null. For GPT-4, there is no clear effect of temperature on $\beta$. With memory 1, we cannot estimate $\beta$ because forecasts remain far from prices, preventing the initial learning phase from concluding (see Section \ref{sec:alignment} and \citealt{heemeijer2009price}).\footnote{Supplementary Figure \ref{fig:fig4supp} shows results obtained without removing the initial learning phase. The findings are broadly consistent with Figure \ref{fig:table_overview}.}  
In positive feedback markets, GPT-3.5 produces $\beta$ estimates that exceed those of human subjects, whereas GPT-4 yields estimates that closely match human behavior. For GPT-3.5, increasing memory and temperature appears to amplify $\beta$ estimates, whereas for GPT-4, this effect is not evident.  

These results suggest that while LLM agents replicate key aspects of human forecasting behavior--such as trend following in positive feedback markets and lack of trend-following in negative feedback markets--they also display notable differences. Memory plays a critical role in shaping LLM behavior, with higher memory settings leading to more human-like convergence patterns. Temperature also influences agent behavior, but its effects are less consistent. These findings highlight the importance of model parameterization when using LLMs to simulate human economic decision-making.

\subsection{LLM reasoning and narratives}
\label{sec:results_textual}

\begin{figure}[!h]
    \centering
    \includegraphics[width=1\textwidth]{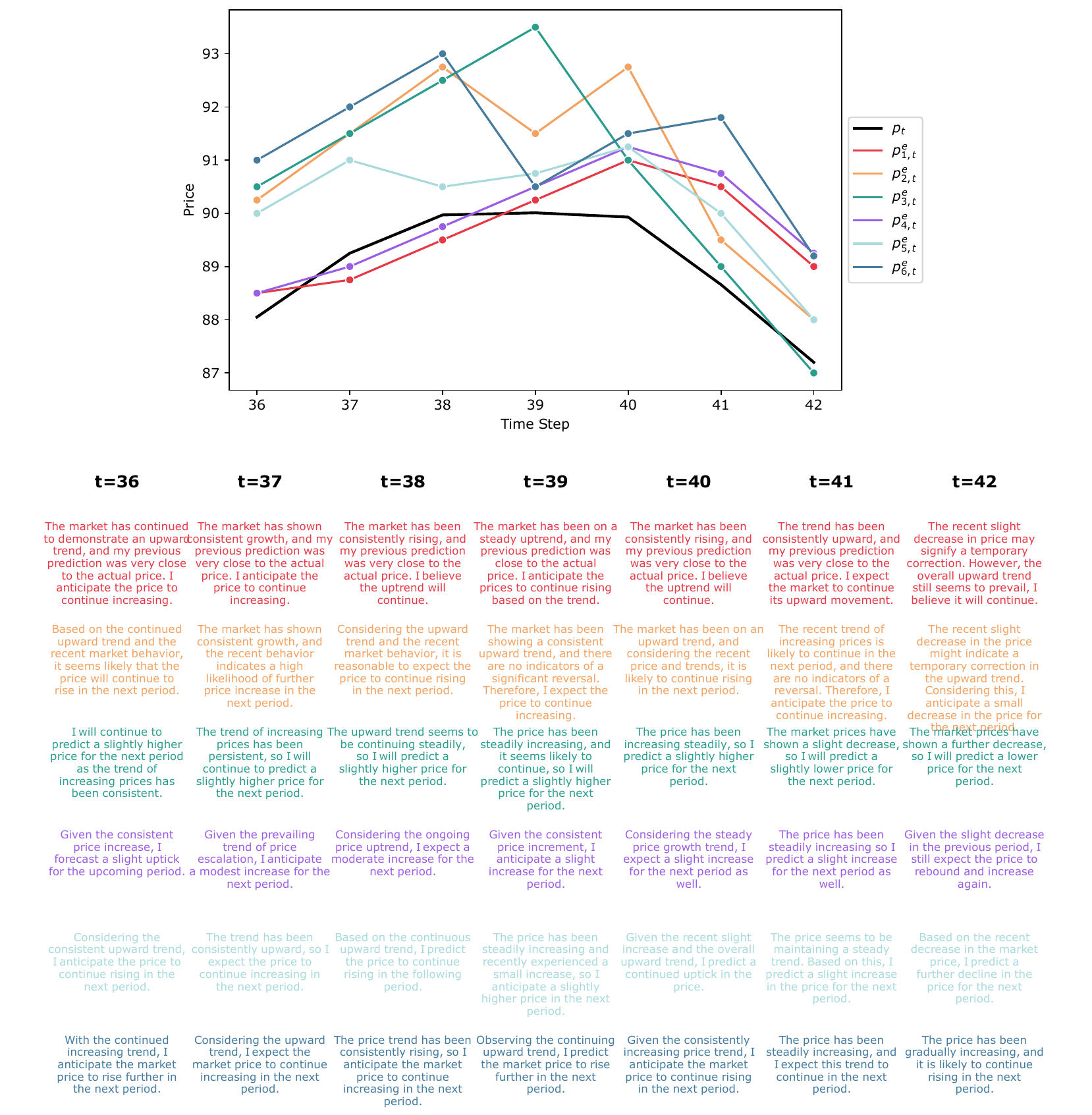}
    \caption{\textbf{Narratives of GPT-3.5 around the bursting of the bubble.} We zoom in on the left-most panel of GPT-3.5 in positive feedback markets in Figure \ref{fig:grid_human_ai}. For clarity, we use different colors for each individual forecast, and report the reasoning of the LLM at each step using the same color. The black line indicates the market price.   }
    \label{fig:text}
\end{figure}

Finally, to better understand LLM predictions, we analyze the reasoning text (i.e. narratives) they provide when making the forecast. Since this information is inherently high-dimensional, we focus on a specific episode where reasoning may be particularly revealing. Figure \ref{fig:text} presents predictions and prices around the bursting of the bubble shown in Figure \ref{fig:grid_human_ai} with GPT-3.5, alongside the narratives provided by the agents.  

A key observation from the narratives is that LLM agents consistently reason as trend followers. For example, at time step 36, agent 1 writes, ``The market has continued to demonstrate an upward trend [...]. I anticipate the price to continue increasing.'' Similarly, agent 2 states, ``Based on the continued upward trend and the recent market behavior, it seems likely that the price will continue to rise in the next period.'' This finding supports the analysis in Section \ref{sec:results_behavioral_parameters}, which identifies LLM agents as trend followers in positive feedback markets.  

Beyond confirming trend-following behavior, the narratives also reveals that expectations shift when prices slow down and eventually decline. From time step 36 to 37, all agents forecast higher prices, driving the market price upward. However, at time step 38, agent 5 predicts a price above the current market level but below its previous forecast, thereby slowing price growth.  

The turning point occurs at time step 39, when agent 6 forecasts a price 2.5 EUR lower than in the previous period, halting price growth entirely. Interestingly, this shift is not reflected in agent 6’s reasoning, as it writes, ``Observing the continuing upward trend, I predict the market price to rise further in the next period.'' This discrepancy suggests that the agent's numerical prediction is influenced by noise, consistent with the high temperature setting used in the experiment.  

As prices stabilize, agents remain generally optimistic, as indicated by their reasoning at $t=41$, yet their numerical predictions begin to decline. The only exception is agent 3, who had been lowering its forecasts since $t=40$ and justifies this by writing, ``The market prices have shown a slight decrease, so I will predict a slightly lower price for the next period.''  

As the market continues to decline, agents adjust their expectations in a highly heterogeneous manner. At time step 42, agent 4 states, ``Given the slight decrease in the previous period, I still expect the price to rebound and increase again,'' while agent 5 writes, ``Based on the recent decrease in the market price, I predict a further decline in the price for the next period.'' Despite these individual differences, all agents eventually align with the downward trend, which persists until the end of the simulation.  

The qualitative analysis of LLM-generated narratives provides strong evidence that GPT-3.5 agents reason as trend followers, reinforcing the quantitative findings from Section \ref{sec:results_behavioral_parameters}. However, the narrative also reveals discrepancies between reasoning and numerical predictions, particularly around turning points, where noise plays a significant role in decision-making. While agents initially maintain bullish expectations even as prices flatten, they eventually adapt to the downward trend, albeit with considerable heterogeneity in individual responses. These findings highlight both the consistency and the limitations of LLM reasoning in market environments.

\section{Discussion}
\label{sec:discussion}

In this study, we explored the potential for LLMs to replicate human behavior in laboratory experiments within economic markets. Specifically, LLM systems replicated the differences between positive and negative feedback markets concerning trend following. However, at a detailed level, differences emerged, such as LLMs being less naive and more obstinate in negative markets. One important finding is that while some context is important for replicating human behavior, it is not necessary to provide the full context window; memory of 3 to 5 previous messages may suffice. This finding implies that simulations may be computationally feasible even for long periods of time.

Our results contribute to the literature on the extent to which LLMs can reproduce human behavior. Similar to \cite{chen2023emergence}, we found some alignment, though with less heterogeneity.  Notably, LLM markets do not follow rational expectations but exhibit a degree of bounded rationality. Compared to other studies \citep{zhaocompeteai,li2024econagent} that have attempted to simulate economic behavior beyond individual actors, our work provides promising results by comparing not only stylized facts (e.g., convergence to the fundamental price) but also by measuring convergence time and forecaster strategies in detail.

LLMs are relatively new, and there is much room for further research. To increase heterogeneity, future work could explore incorporating demographics or political values, as \cite{argyle2023out} did. Another approach could involve integrating the literature on investment strategies and personality traits (e.g., risk-seeking versus risk-averse behaviors). One could also exploring more complex dynamics, such as bubble formation, adaptive learning \citep{hommes2008expectations}. These studies may also increase understanding of LLM intelligence, particularly in relation to theory of mind \citep{guo2023suspicion,strachan2024testing}.  Key challenges include determining  \textit{which human} economic behavior LLMs represent, considering that the literature suggests LLMs may reflect more Western and left-leaning political views \citep{rozado2024political,atari2023humans} and developing robust validation frameworks \citep{larooij2025llm}.

Achieving accurate modeling of human behavior in silico \citep{horton2023large} would have significant implications for the economics literature. This development could enable a general-purpose bounded rationality framework that can be applied to model decisions of a wide variety of economic agents \citep{pangallo2024datadriven}. While we are only at the beginning, the rapid advances in AI suggest that this could mark the start of a revolution in economics, similar to the impact of behavioral experiments.

\section{Acknowledgments}
The authors acknowledge funding from OpenAI for credits used to run the experiments. We are also grateful to Pamela Mishkin for her insightful comments and for engaging in valuable brainstorming discussions during the early stages of this project.
Opinions expressed in this paper are those of the authors and do not necessarily
reflect those of the Bank of Canada or its staff.

%\bibliographystyle{agsm}
% \bibliographystyle{plain}
%\bibliography{references}

\newpage
\appendix

\label{sec:appendix}

\section{Experimental instructions}
\label{apx:instructions}

\subsection*{Experimental instructions for negative feedback treatment}

\subsubsection*{B.1.1. Experimental instructions}
The shape of the artificial market used by the experiment, and the role you will have in it, will be explained in the text below. Read these instructions carefully. They continue on the backside of this sheet of paper.

\subsubsection*{B.1.2. General information}
You are an advisor of an importer who is active on a market for a certain product. In each time period, the importer needs a good prediction of the price of the product. Furthermore, the price should be predicted one period ahead, since importing the good takes some time. As the advisor of the importer, you will predict the price $P(t)$ of the product during 50 successive time periods. Your earnings during the experiment will depend on the accuracy of your predictions. The smaller your prediction errors, the greater your earnings.

\subsubsection*{B.1.3. About the market}
The price of the product will be determined by the law of supply and demand. The size of demand is dependent on the price. If the price goes up, demand will go down. The supply on the market is determined by the importers of the product. Higher price predictions make an importer import a higher quantity, increasing supply. There are several large importers active on this market, and each of them is advised by a participant of this experiment. Total supply is largely determined by the sum of the individual supplies of these importers. Besides the large importers, a number of small importers are active on the market, creating small fluctuations in total supply.

\subsubsection*{B.1.4. About the price}
The price is determined as follows. If total demand is larger than total supply, the price will rise. Conversely, if total supply is larger than total demand, the price will fall.

\subsubsection*{B.1.5. About predicting the price}
The only task of the advisors in this experiment is to predict the market price $P(t)$ in each time period as accurately as possible. The price (and your prediction) can never become negative and always lies between 0 and 100 euros in the first period. The price and the prediction in period 2 through 50 is only required to be positive. The price will be predicted one period ahead. At the beginning of the experiment, you are asked to give a prediction for period 1, $V(1)$. When all participants have submitted their predictions for the first period, the market price $P(1)$ for this period will be made public. Based on the prediction error in period 1, $P(1) - V(1)$, your earnings in the first period will be calculated. Subsequently, you are asked to enter your prediction for period 2, $V(2)$. When all participants have submitted their prediction for the second period, the market price for that period, $P(2)$, will be made public and your earnings will be calculated, and so on, for 50 consecutive periods. The information you have to form a prediction at period $t$ consists of: All market prices up to time period $t-1$: $\{P(t-1), P(t-2), \dots, P(1)\}$;  All your predictions up until time period $t-1$: $\{V(t-1), V(t-2), \dots, V(1)\}$; Your total earnings at time period $t-1$.

\subsubsection*{B.1.6. About the earnings}
Your earnings depend only on the accuracy of your predictions. The better you predict the price in each period, the higher will be your total earnings. The attached table lists all possible earnings.

When you are done reading the experimental instructions, you may continue reading the computer instructions, which have been placed on your desk as well.

\subsection*{Experimental instructions for positive feedback treatment}

\subsubsection*{B.2.1. Experimental instructions}
The shape of the artificial market used by the experiment, and the role you will have in it, will be explained in the text below. Read these instructions carefully. They continue on the backside of this sheet of paper.

\subsubsection*{B.2.2. General information}
You are an advisor of a trader who is active on a market for a certain product. In each time period, the trader needs to decide how many units of the product he will buy, intending to sell them again the next period. To take an optimal decision, the trader requires a good prediction of the market price in the next time period. As the advisor of the trader, you will predict the price $P(t)$ of the product during 50 successive time periods. Your earnings during the experiment will depend on the accuracy of your predictions. The smaller your prediction errors, the greater your earnings.

\subsubsection*{B.2.3. About the market}
The price of the product will be determined by the law of supply and demand. Supply and demand on the market are determined by the traders of the product. Higher price predictions make a trader demand a higher quantity. A high price prediction makes the trader willing to buy the product; a low price prediction makes him willing to sell it. There are several large traders active on this market, and each of them is advised by a participant of this experiment. Total supply is largely determined by the sum of the individual supplies and demands of these traders. Besides the large traders, a number of small traders are active on the market, creating small fluctuations in total supply and demand.

\subsubsection*{B.2.4. About the price}
The price is determined as follows. If total demand is larger than total supply, the price will rise. Conversely, if total supply is larger than total demand, the price will fall.

\subsubsection*{B.2.5. About predicting the price}
The only task of the advisors in this experiment is to predict the market price $P(t)$ in each time period as accurately as possible. The price (and your prediction) can never become negative and always lies between 0 and 100 euros in the first period. The price and the prediction in period 2 through 50 is only required to be positive. The price will be predicted one period ahead. At the beginning of the experiment, you are asked to give a prediction for period 1, $V(1)$. When all participants have submitted their predictions for the first period, the market price $P(1)$ for this period will be made public. Based on the prediction error in period 1, $P(1) - V(1)$, your earnings in the first period will be calculated. Subsequently, you are asked to enter your prediction for period 2, $V(2)$. When all participants have submitted their prediction for the second period, the market price for that period, $P(2)$, will be made public and your earnings will be calculated, and so on, for 50 consecutive periods. The information you have to form a prediction at period $t$ consists of: All market prices up to time period $t-1$: $\{P(t-1), P(t-2), \dots, P(1)\}$; All your predictions up until time period $t-1$: $\{V(t-1), V(t-2), \dots, V(1)\}$; Your total earnings at time period $t-1$.

\subsubsection*{B.2.6. About the earnings}
Your earnings depend only on the accuracy of your predictions. The better you predict the price in each period, the higher will be your total earnings. The attached table lists all possible earnings.

When you are done reading the experimental instructions, you may continue reading the computer instructions, which have been placed on your desk as well.

\section{LLM instructions}
\label{apx:llm_instructions}

\begin{lstlisting}
    [
    {
        "role": "system",
        "content": "General instructions. You are an advisor of an importer who is active on a market for a certain product. In each time period the importer needs a good prediction of the price of the product. Furthermore, the price should be predicted one period ahead, since importing the good takes some time. As the advisor of the importer you will predict the price P(t) of the product during 50 successive time periods. Your earnings during the experiment will depend on the accuracy of your predictions. The smaller your prediction errors, the greater your earnings."
    },
    {
        "role": "system",
        "content": "About the market. The price of the product will be determined by the law of supply and demand. The size of demand is dependent on the price. If the price goes up, demand will go down. The supply on the market is determined by the importers of the product. Higher price predictions make an importer import a higher quantity, increasing supply. There are several large importers active on this market and each of them has an advisor like you. Total supply is largely determined by the sum of the individual supplies of these importers. Besides the large importers, a number of small importers is active on the market, creating small fluctuations in total supply."
    },
    {
        "role": "system",
        "content": "About the price. The price is determined as follows. If total demand is larger than total supply, the price will rise. Conversely, if total supply is larger than total demand, the price will fall."
    },
    {
        "role": "system",
        "content": "About predicting the price. The only task of the advisors is to predict the market price P(t) in each time period as accurately as possible. The price (and your prediction) can never become negative and always lies between 0 and 100 euros in the first period. The price and the prediction in period 2 through 50 is only required to be positive. The price will be predicted one period ahead. At the beginning of the experiment you are asked to give a prediction for period 1, V(1). When all advisors have submitted their predictions for the first period, the market price P(1) for this period will be made public. Based on the prediction error in period 1, P(1) - V(1), your earnings in the first period will be calculated. Subsequently, you are asked to enter your prediction for period 2, V(2). When all advisors have submitted their prediction for the second period, the market price for that period, P(2), will be made public and your earnings will be calculated, and so on, for 50 consecutive periods."
    },
    {
        "role": "system",
        "content": "About the earnings. Your earnings depend only on the accuracy of your predictions. The better you predict the price in each period, the higher will be your total earnings."
    },
    {
        "role": "system",
        "content": "Your prediction can have two decimal numbers, for example 30.75. The available information for predicting the price of the product in period t consists of: All product prices from the past up to period t-1; Your predictions up to period t-1; Your earnings until then"
    },
     {
      "role": "system",
     "content":  "From the second period onwards, you will get the following data: ```market prices: [P(t-1), P(t-2), ..., P(1)]; your predictions: [V(t-1), V(t-2), ..., V(1)]; Total earnings: total accumulated earnings until time t-1```"
    },
    {
        "role": "system",
        "content":  "Response format:  Your response should be exclusively in JSON format with two keys: 'reasoning' where you explain your rationale and method for predicting in 30-50 words, and 'predictedValue', the numeric value of your predicted market price. Nothing outside the JSON format should be written"    
    }]
\end{lstlisting}

\section{Additional results}
\label{apx:timeseries_all}

\begin{figure}[H]
    \centering
    \includegraphics[width=0.8\textwidth]{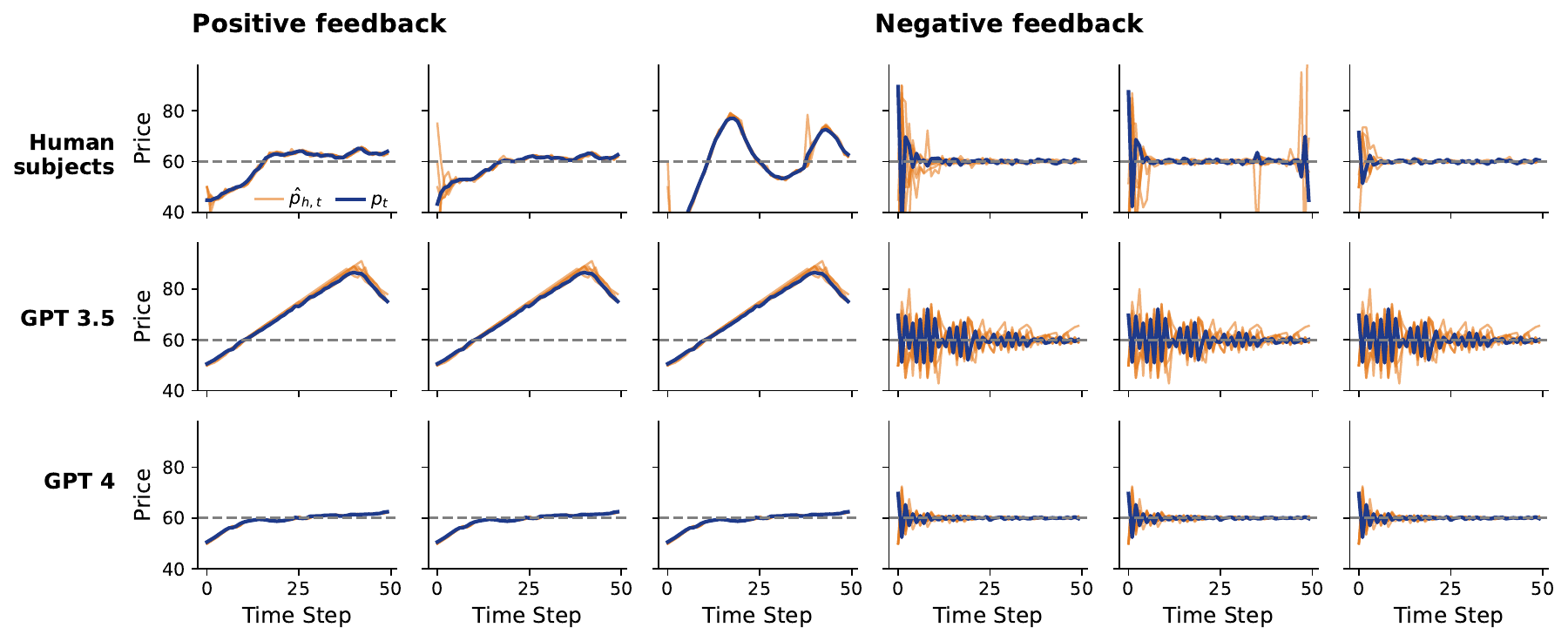}
    \caption{\textbf{Market dynamics for human subjects and LLM agents.} We compare three different experiments for both positive and negative feedback markets, different from the ones shown in Figure \ref{fig:grid_human_ai}. For each experiment, we show both the time series of realized market price (blue) and all agents' expectations (orange). For this illustration of LLM behavior, we select a specific combination of memory (5) and temperature (0.7) that is different from the one in Figure \ref{fig:grid_human_ai}, but that shows qualitatively similar behavior.}
    \label{fig:Gridplot_human_models_supp}
\end{figure}

\begin{figure}[H]
    \centering
    \includegraphics[width=1\textwidth]{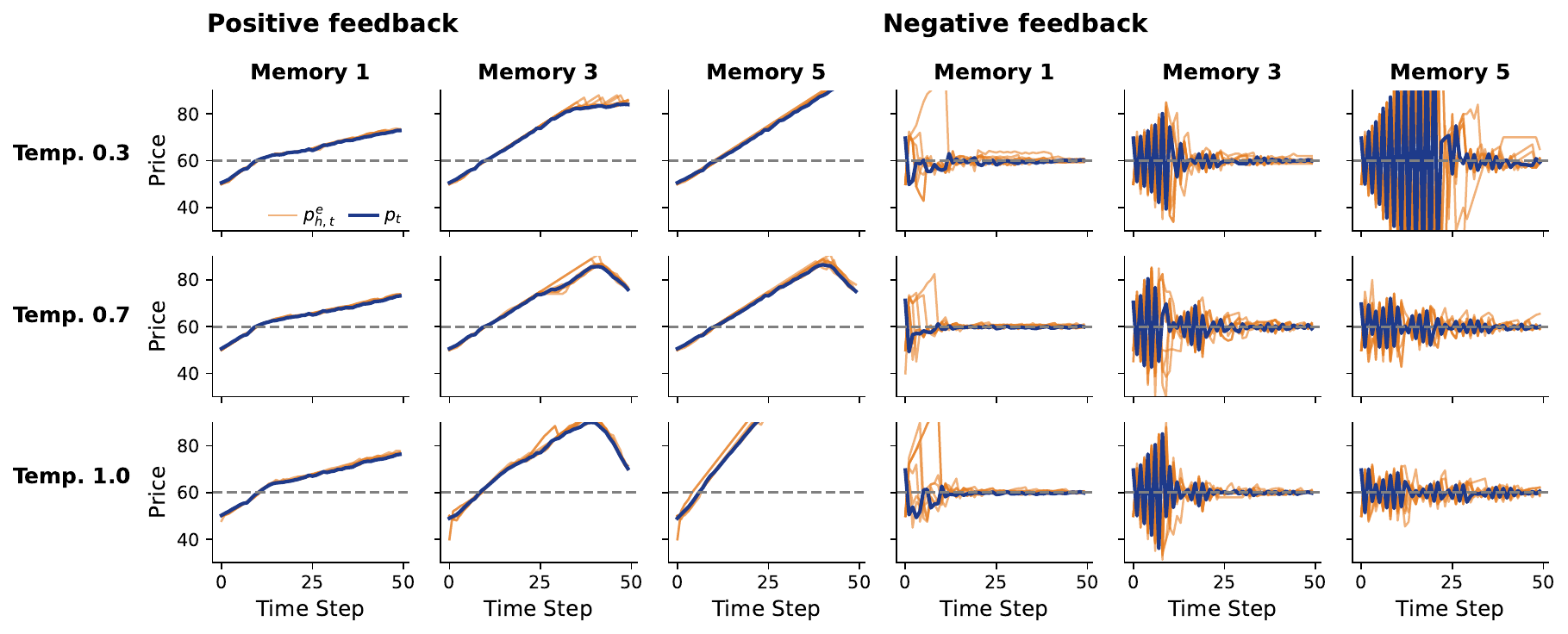}
    \caption{\textbf{Market dynamics for GPT-3.5 agents under different parameters.} We compare all nine combinations of memory (1, 3, 5) and temperature (0.3, 0.7, 1.0), for both positive and negative feedback markets. For each experiment, we show both the time series of realized market price (blue) and all agents' expectations (orange).}
    \label{fig:grid_gpt3}
\end{figure}

\begin{figure}[H]
    \centering
    \includegraphics[width=1\textwidth]{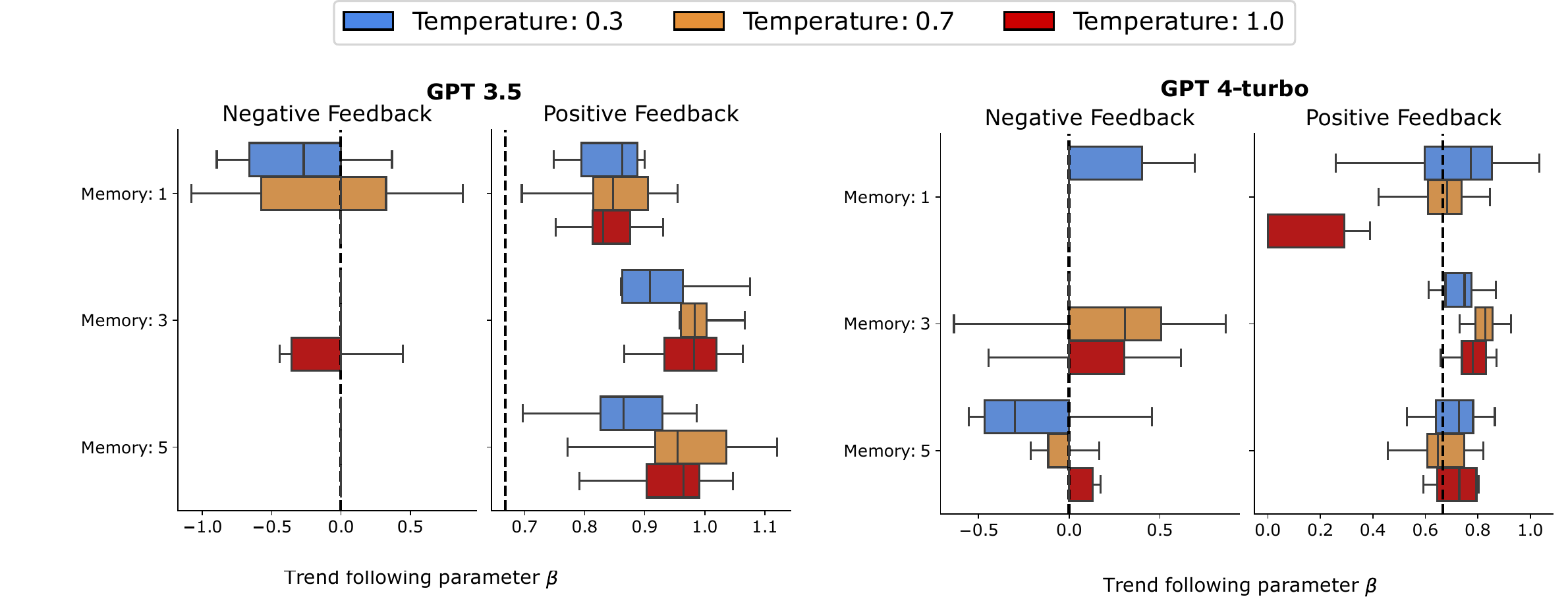}
    \caption{\textbf{Trend following behavior across different memory and temperature parameters.} Each box-plot shows the median, interquartile range, and whiskers of all the estimated $\beta$ values for LLM agents. Dashed vertical lines indicate the average $\beta$ values for human participants, namely 0 for negative feedback markets and 0.67 for positive feedback markets. Compared to Figure \ref{fig:table_overview}, here we do not remove the initial learning phase.}
    \label{fig:fig4supp}
\end{figure}

\end{document}